\documentclass[iop]{emulateapj}
\usepackage{natbib}
\usepackage{txfonts}
\usepackage{graphicx}
\usepackage{wasysym}

\newcommand{\mub}{$\mu_{B}$}

\newcommand{\bmv}{B$-$V}

\begin{document}

\title{DEEP IMAGING OF M51: A NEW VIEW OF THE WHIRLPOOL'S EXTENDED
  TIDAL DEBRIS}

\shortauthors{Watkins et al.}
\shorttitle{Extended tidal debris of M51}

\author{Aaron E. Watkins\altaffilmark{1}, 
  J. Christopher Mihos\altaffilmark{1},
  Paul Harding\altaffilmark{1}}

\altaffiltext{1}{Department of Astronomy, Case Western Reserve
  University, Cleveland, OH 44106, USA}
  
\begin{abstract}

We present deep, wide-field imaging of the M51 system using CWRU's
  Burrell Schmidt telescope at KPNO to study the faint tidal features
that constrain its interaction history. Our images trace M51's tidal
morphology down to a limiting surface brightness of $\mu_{B,lim}\sim
$30 mag arcsec$^{-2}$, and provide accurate colors ($\sigma_{\bv} <
0.1$) down to $\mu_B\sim 28$. We identify two new tidal streams in the
system (the South and Northeast Plumes) with surface brightnesses of
$\mu_B =29$ and luminosities of $\sim 10^6 L_{\astrosun,B}$. While the
Northeast Plume may be a faint outer extension of the tidal ``crown''
north of NGC 5195 (M51b), the South Plume has no analogue in any
existing M51 simulation and may represent a distinct tidal stream or
disrupted dwarf galaxy. We also trace the extremely diffuse Northwest
Plume out to a total extent of 20\arcmin\ (43 kpc) from NGC 5194
(M51a), and show it to be physically distinct from the overlapping
bright tidal streams from M51b. The Northwest Plume's morphology and
red color ($\bv=0.8$) instead argue that it originated from tidal
stripping of M51a's extreme outer disk. Finally, we confirm the strong
segregation of gas and stars in the Southeast Tail, and do not detect
any diffuse stellar component in the \ion{H}{1} portion of the
tail. Extant simulations of M51 have difficulty matching both the
wealth of tidal structure in the system and the lack of stars in the
\ion{H}{1} tail, motivating new modeling campaigns to study the
dynamical evolution of this classic interacting system.

\end{abstract}


\newpage

\section{Introduction}

Galaxy interactions form the basis of the current hierarchical
accretion paradigm for galaxy evolution, and have been linked to the
formation of bars, spiral structure, and induced star formation.
However, all these features may also result from purely secular
processess \citep[see the review by][]{athanassoula10}, entangling the
various evolutionary effects in individual systems. Understanding
interaction histories of nearby galaxies, from which much of our
detailed understanding of physical processes comes, is thus
particularly important.

The galaxy pair M51 is one of the best known nearby interacting
systems, comprised of the grand-design spiral NGC 5194 (M51a) and its
early-type SB0 companion NGC 5195 (M51b). Deep imaging in the optical
and infrared \citep[e.g.][]{burkhead78, smith90, mutchler05} coupled
with kinematic studies in optical \citep[e.g.][]{tully74, durrell03}
and radio \citep[e.g.][]{appleton86, rots90}, have revealed extended
tidal streams and complex kinematics, including an apparently
counter-rotating tail of gas on its southern end \citep{rots90}.

M51's proximity \citep[7.5 Mpc;][]{ciardullo02, bose14} and
well-defined spiral structure has made it a frequent target for
studies of the interplay between spiral dynamics, star formation, and
the atomic and molecular ISM \citep[e.g.][]{calzetti05, hughes13}. If
much of the star formation present in the system is
interaction-induced, models become an important means of constraining
the physical mechanisms and timescales involved, and hence, for
example, the calibration of star formation tracers in galaxies
\citep[e.g.][]{calzetti05} and determination of generalized star
formation laws \citep[e.g.][]{kennicutt07}.

This has made the pair a favorite subject of interaction models
\citep[e.g.][and many others]{toomre72, howard90, salo00, durrell03,
  dobbs10}, constrained by the morphology and kinematics of the
observed tidal features. These modeling campaigns have led to two
classes of models for the system: one involving a single fly-by
passage \citep{toomre72, hernquist90, durrell03}, and a second
involving multiple passes on a more tightly bound orbit \citep{salo00,
  dobbs10}. Discriminating between these scenarios can constrain both
the timescale and strength of the perturbation that set up the
grand-design spiral and induced star formation in the M51 system.
  
\begin{figure*}
  \centering
  \includegraphics[scale=0.32]{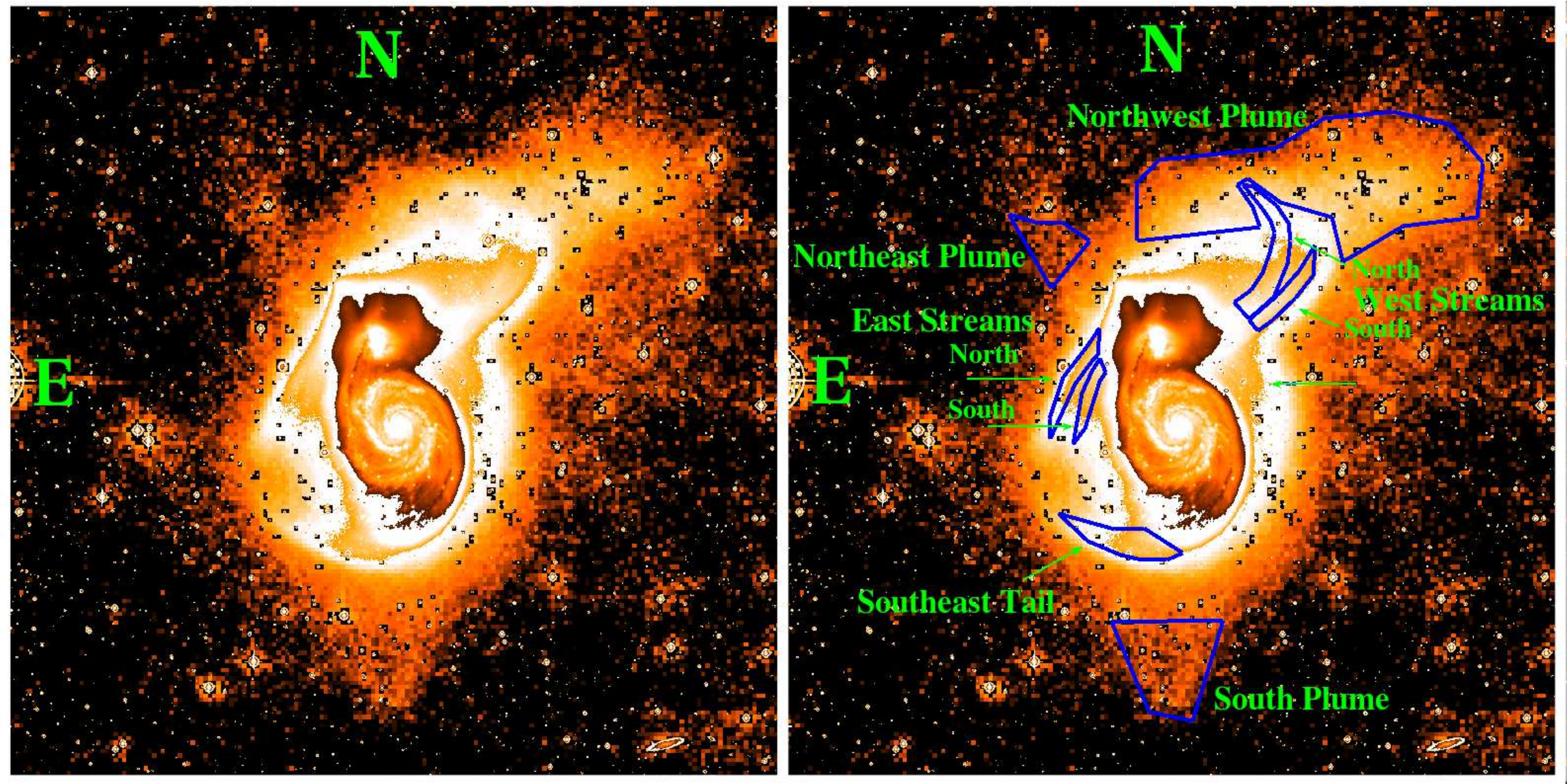}
  \caption{{\bf Left:} a subset of our B band mosaic, rescaled in
    intensity for different surface brightness ranges: \mub$<24.6$,
    $24.6<$\mub$<26.5$, and \mub$>26.5$. At the lowest surface
    brightnesses, the image has been median binned to enhance faint
    structures.  {\bf Right:} as left, marking regions detailed in
    Table \ref{table:m51} and discussed in the text.  Each image is
    39\arcmin\ (84 kpc) on a side.
    \label{fig:regions}}
\end{figure*}

Deeper observations can prove useful in locating additional
interaction signatures to further constrain the models. Particularly
in M51, in which both galaxies have vastly different star formation
rates \citep{lee11, lee12}, broadband colors may be used to help
determine the origin of many interaction signatures, and add
additional constraints from stellar populations. Combining these
observations with the kinematics and spatial distribution of the gas
in the system can also greatly narrow down the available parameter
space for the dynamical models. In this {\it Letter}, we present deep
multiband optical imaging of M51, taken as part of a survey of low
surface brightness features of nearby galaxies, using the 0.6/0.9m
  CWRU Burrell Schmidt Telescope at Kitt Peak National Observatory
\citep{watkins14}. We report on the morphology and colors of both
known and newly-discovered tidal features which will help constrain
the dynamical evolution of the M51 system.

\begin{deluxetable*}{l c c c c c c}
\tabletypesize{\small}
\tablecaption{Photometry of Tidal Debris\label{table:m51}}
\tablecolumns{7}
\tablehead{
\colhead{Region} & \colhead{\mub$_{,peak}$} & \colhead{$M_{B}$} & 
\colhead{\bv} & \colhead{$10^{6} L_{\ensuremath{\odot},B}$} &
\colhead{Origin} & \colhead{Gas Content}}
\startdata
\\[0.01cm]
West Stream -- North & 25.43 & -14.3 & 0.75 & 45 & M51b & Gas Poor \\[0.1cm]
West Stream -- South & 25.46 & -14.2 & 0.73 & 39 & M51b & Gas Poor  \\[0.1cm]
East Stream -- North & 25.78 & -14.1 & 0.73 & 39 & M51b & Gas Poor  \\[0.1cm]
East Stream -- South & 25.63 & -13.9 & 0.72 & 30 & M51b & Gas Poor  \\[0.1cm]
Southeast Tail & 25.80 & -14.8 & 0.64 & 68 & M51a & Gas Rich, offset\\[0.1cm]
Northwest Plume & 27.59 & -14.5 & 0.81 & 101 & M51a? & Patchy? \\[0.1cm]
Northeast Plume & 29.0$\pm$0.3 & -11.2 & --- & 4 & M51a? & Gas Poor \\[0.1cm]
South Plume & 29.2$\pm$0.7 & -11.8 & --- & 8 & Unknown & Gas Poor 
\enddata
\tablecomments{
 Absolute photometric errors are $\sigma_B \sim 0.04$ and
   $\sigma_{B\!-\!V} \sim 0.05$.}
\end{deluxetable*}
    
\section{Observations and Analysis}

We give a brief account of our observation and data reduction
procedures here; more details can be found in \citet{watkins14} and
references therein. We observed M51 in spring 2010 and spring 2012,
with a setup identical to that described in \citet{watkins14}.  We
observed only on moonless photometric nights, in two filters: modified
Johnson \emph{B} (31$\times$1200s exposures) and Washington \emph{M}
(39$\times$900s exposures). We generated dark sky flats using a
comparable number of offset night sky images in each filter
\citep{watkins14}. Photometric conversion to standard Johnson \emph{B}
and \emph{V} used observations of \citet{landolt92} standard fields to
determine color terms, and on-frame SDSS \citep[DR8;][]{aihara11}
stars to derive direct photometric zeropoints.

\begin{figure*}
  \centering
  \includegraphics[scale=0.8]{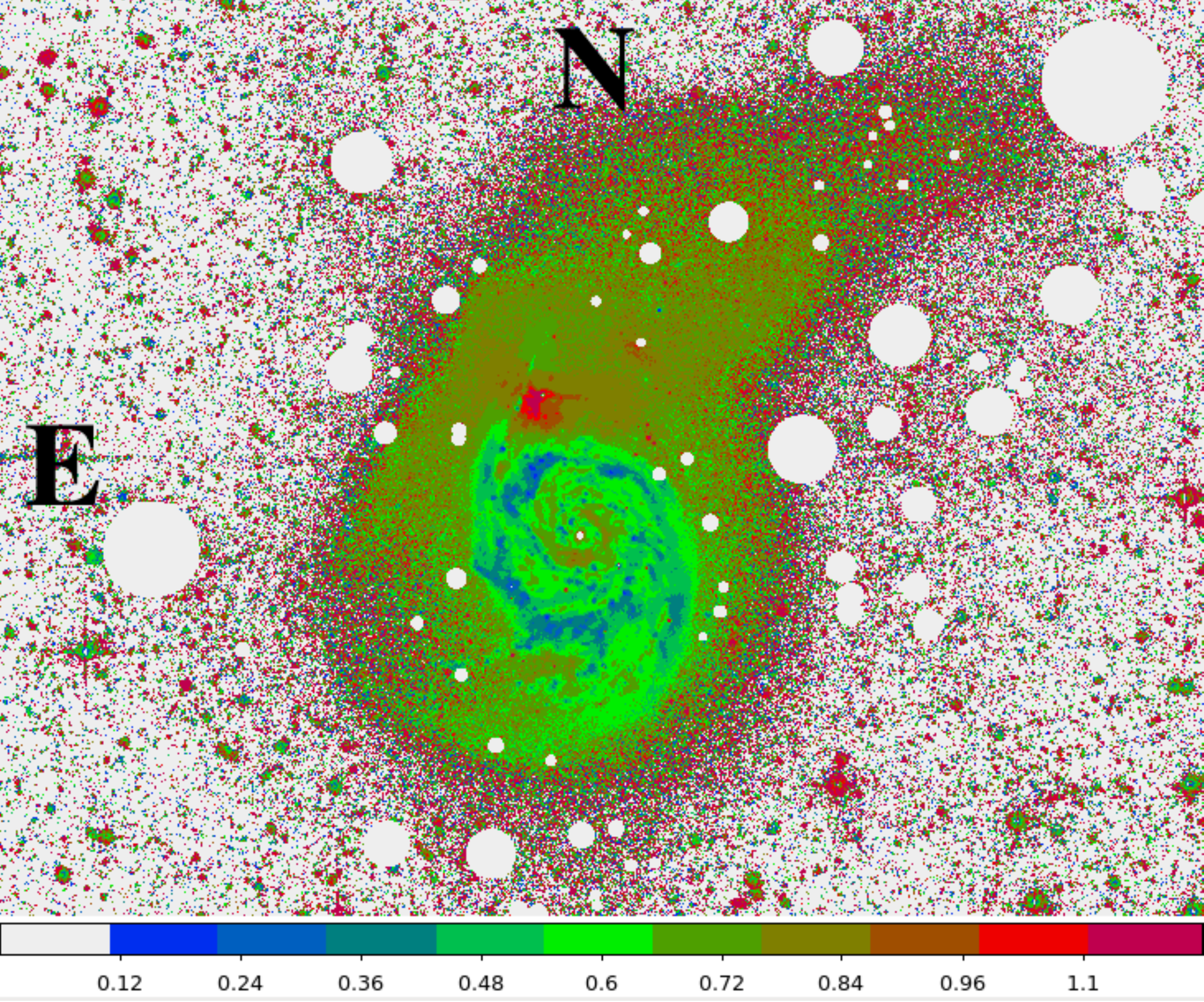}
  \caption{\bmv\ colormap of the M51 system. The image spans
    36\arcmin~$\times$~27\arcmin \ (78 kpc~$\times$~58 kpc).
    \label{fig:color}}
\end{figure*}

Data reduction consisted of standard reduction procedures and modeling
and subtraction of reflections and extended wings of bright stars
\citep[see][]{slater09}.  Following sky subtraction, each image was
spatially registered and scaled to a common photometric zeropoint
before median combining all images into the final \emph{B} and
\emph{V} mosaics. We then masked the final mosaics of bright discrete
sources and spatially rebinned and median filtered them in 9$\times$9
pixel blocks to enhance faint, diffuse features. Our final \emph{B}
image and our colormap of M51 are shown in Figures~\ref{fig:regions}
and \ref{fig:color}, respectively.  Our limiting magnitude is
\mub$\sim30$.

In the quantitative photometric analysis that follows, we use
polygonal apertures to define regions of interest in M51 (see Figure
\ref{fig:regions} and Table \ref{table:m51}). In these regions,
contamination from background sources can be significant, so we
subtract off a local background flux \citep[see][for
  details]{rudick10, mihos13}.  At extremely low surface brightnesses,
this background contamination dominates the uncertainty, and is
non-Gaussian.  At surface brightnesses of \mub$<27.5$, the relative
uncertainties are small, but at extremely low surface brightness
(\mub$>28.5$) they become significant and preclude measurement of
meaningful colors. All values have also been corrected for the local
foreground extinction \citep[$A_{B} = 0.126$, $E(\bv) =
  0.031$;][]{schlafly11}.

In searching for tidal features in deep surface photometry, an
additional source of contamination is scattered light from foreground
Galactic dust, which can mimic diffuse structures
\citep[see][]{rudick10}. However, such dust radiates in the mid- and
far-infrared, and a comparison of our images with the WISE 12$\mu$m
and IRIS 100$\mu$m images \citep[][respectively]{meisner14, miville05}
shows no evidence for strong dust contamination cospatial with the
diffuse features we identify in Figure \ref{fig:regions} (to
  limits of $3.58 M\!Jy~sr^{-1}$/$1.11 M\!Jy~sr^{-1}$ in WISE/IRIS).

\section{Results}

\subsection{Morphology and color of tidal features}

\begin{figure*}
  \centering
  \includegraphics[scale=0.4]{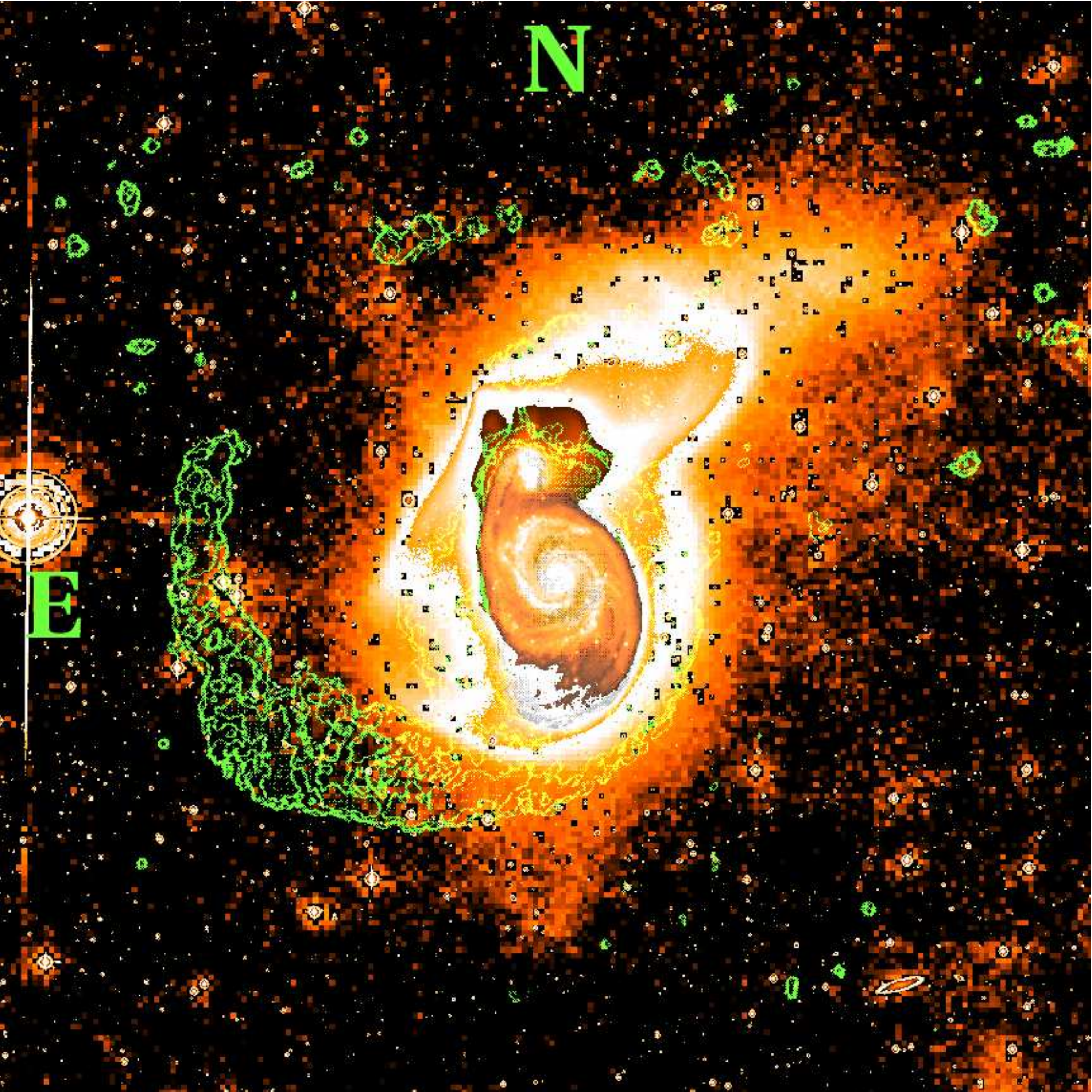}
  \caption{Comparison between our optical image and (in green
    contours) the \ion{H}{1} data of \citet{rots90}.
    \label{fig:gascomp}}
\end{figure*}

An examination of Figure~\ref{fig:regions} shows a wealth of tidal
debris around M51. At higher surface brightness (\mub $<26.5$) we see
many well-known features, including M51's southern tidal tail, the
east and west tidal streams emanating from M51b's disk, and the
three-pronged structure (the ``crown'') just north of M51b. At
extremely low surface brightness, we trace the extended Northwest
Plume, and identify two new, extremely faint features---the Northeast
and South Plumes. We also show in Figure \ref{fig:gascomp} a
comparison between our optical image and the \ion{H}{1} data from
\citet{rots90}.
 
The most extended feature is the Northwest Plume. Originally detected
by \citet{burkhead78}, its tip lies 20\arcmin \ (43 kpc) from the
center of M51a. While it partially blends with the brighter West
Streams extending from M51b, a variety of arguments indicate it is a
separate feature.  First, the Plume is morphologically distinct---it
extends almost directly west, while the northern West Stream extends
first northwest and then curves toward the northeast (Figure
\ref{fig:regions}). The Northwest Plume is also $\sim$0.06 magnitudes
redder\footnote{While this difference is comparable to our {\it
      absolute} photometric uncertainty, {\it relative} color
    differences between adjacent features are constrained to $\sim
    0.01$ mag uncertainty, due to the stability of the background
    corrections.} than the adjacent West Streams (Table
\ref{table:m51}; we note that the plume shows a uniform color from end
to end), implying distinct (albeit similar) stellar
populations. \citet{durrell03} also note bimodality in the PNe
velocities northwest of M51 (although not cospatial with the Plume),
arguing for kinematically distinct populations. Indeed, we note a
possible third component, a slight protrusion just to the northwest of
M51a (marked with an arrow in Figure \ref{fig:regions}) which may be a
stellar stream overlapping M51a's outer disk. The THINGS \ion{H}{1}
map of M51 \citep{walter08} also shows gas extending roughly in this
direction.

There is very little high column density \ion{H}{1} associated with
the Northwest Plume (Figure \ref{fig:gascomp}), but evidence exists
for extremely diffuse, high-velocity gas in the vicinity
\citep[][Pisano private communication]{appleton86, rots90}. Diffuse
\ion{H}{1} associated with the Northwest Plume would also imply an
origin with the gas-rich spiral M51a, as the companion galaxy M51b
appears to be extremely gas-poor.  

Dynamically, the Northwest Plume appears several hundred Myrs old; its
linear extent ($\sim$40 kpc in projection) would require an unrealistic
expansion velocity if, for example, it had formed during the most recent
passage proposed by \citet{salo00} (50 to 100 Myr ago) or
\citet{howard90} (70 Myr ago). At the rotation speed of M51a \citep[210
km/s;][]{appleton86}, a timescale of $\sim$200 Myr is more likely.
We have also used stellar population modeling \citep{kotulla09} to show
the Plume is dominated by old stars. Exponentially declining star
formation histories typical of Sb/Sc galaxies ($\tau=$5--9 Gyr) produce
colors too blue ($B-V\lesssim0.6$) to match the Plume; its red color
demands much older star formation histories ($\tau=$3 Gyr). Similarly,
any subsequent interaction-induced star formation in the Plume must be
weak ($<$5\% of the Plume's stellar mass), else the colors would too
blue by $>$ 0.15 mags even after 500 Myr. As dust extinction appears
negligible due to the low gas column density \citep{appleton86, rots90},
our inference of old stellar populations appears sound.

Partially overlapping with the Northwest Plume, but with contrasting
morphology and color, are the East and West Streams. Both streams are
bifurcated, and all branches have very similar colors ($\bv=0.73$),
which are in turn similar to M51b's average color
\citep[$\bv=0.7$;][]{lee12}. Given this, and the streams' symmetry
around and proximity to M51b, it seems likely that these streams
originated from that galaxy's disk. A variety of processes may give
rise to bifurcated tidal streams, including kinematic caustics
\citep{struck12, smith14} and differences in the kinematic evolution
of the gas and stellar components \citep{mihos01}. In the latter
mechanism, stellar bifurcations would only arise via subsequent star
formation in the displaced gas tail, which would then show younger
stellar populations than the collisionless tail. The lack of
\ion{H}{1} and the uniformly red colors of these streams would
  seem to argue against such a scenario in M51b, and instead
favor collisionless kinematic caustics.

Based on the colors shown in Figure \ref{fig:color}, much of the crown
north of M51b also appears associated with the companion, excepting
the bluer middle prong. \citet{lee12} showed that the latter is
  an extension of M51a's northern tidal arm; indeed, its color is
  similar to that of the Southeast Tail (\bmv=0.64), the blueness of
  which is expected from its origin in M51a's disk. We also note the
incredible sharpness of the ridge formed by the crown's eastern prong,
indicative of a strong caustic.

In the Southeast Tail, the clear separation between the gas and stars
(Figure \ref{fig:gascomp}, noted previously by \citet{howard90} and
\citet{rots90}) is of particular interest. The stellar tail curves
sharply northward and disappears into the tidal detritus near the East
Streams, while the more loosely wrapped \ion{H}{1} tail is much more
extended and contains no diffuse stellar counterpart down to our limit
of \mub$=30.5$. Simulations have shown that such gas/star
offsets in tidal tails can arise from differences in the radial extent
of the two components in the pre-encounter disk \citep{mihos01}, where
the stellar component of a tail consists preferentially of material
with low specific angular momentum and thus ``trails'' the high
angular momentum gaseous tail. While this may explain the tail's lack
of an old stellar population, the lack of either ongoing star
formation \citep{gildepaz07} or a young stellar population is more
challenging, as the compressive nature of tidal interactions is known
to form massive star clusters and tidal dwarf galaxies
\citep{barnes92, jarrett06} in many tidal tails.  The lack of a young
population in M51's tail may indicate a highly inclined encounter,
where compressive forces in the tail are greatly reduced. As shown by
\citet{rots90}, an inclined encounter also explains the tail's
observed counter-rotation, and is reproduced well by interaction
models of \citet{salo00}.

\begin{figure*}
  \centering
  \includegraphics[scale=0.7]{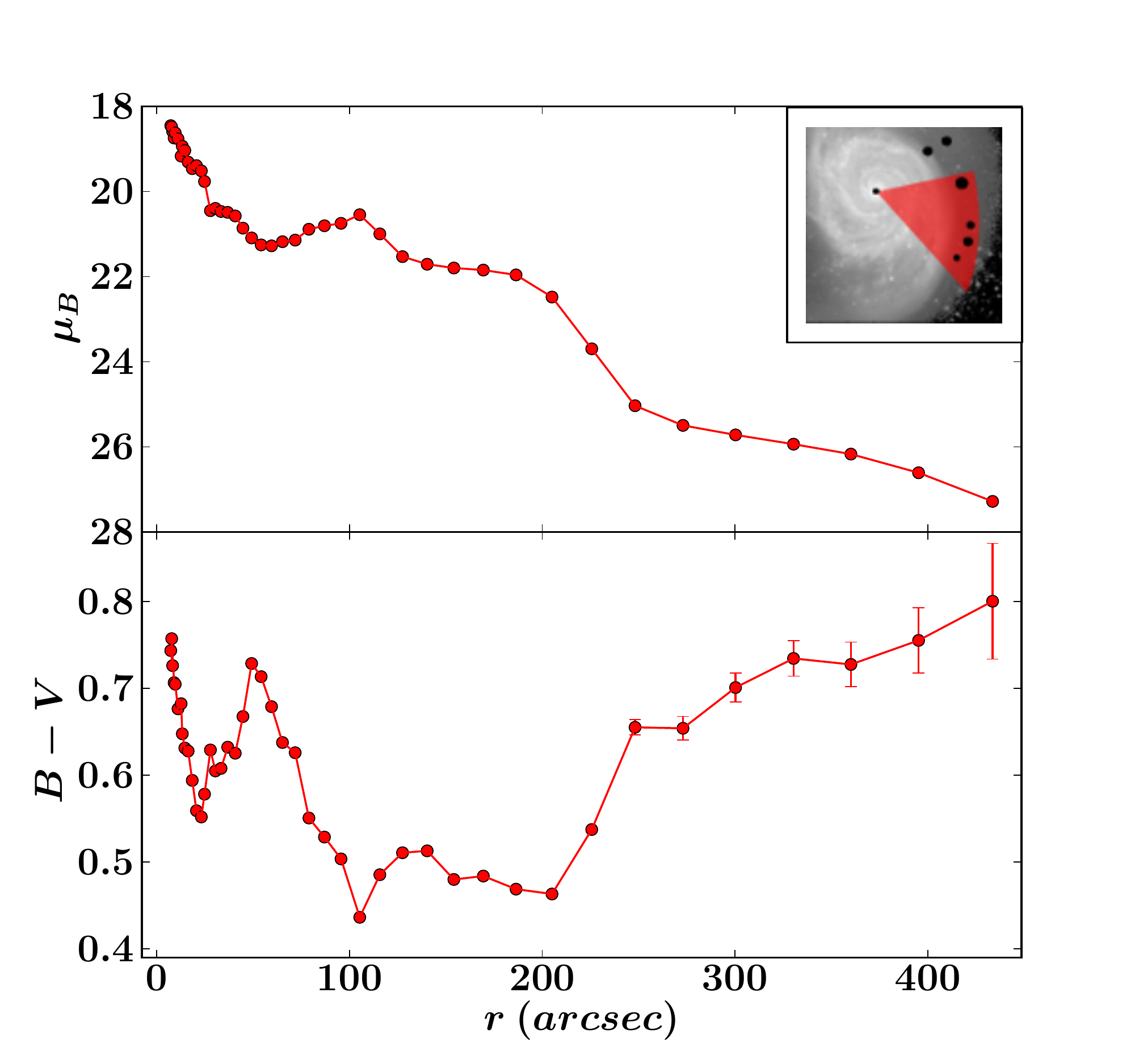}
  \caption{Surface brightness and color profiles along the wedge
  highlighted in the inset, plotted as a function of the average radial
  distance from M51a's center. Error bars in both plots are calculated
  using the background variance, and are often smaller than the point
  size.
  \label{fig:profs}}
\end{figure*} 

Aside from the structure of the tidal features, our data also shed
light on the structure of M51a's outer disk. The colors get
  redder by 0.2--0.4 magnitudes outside of the spiral arm region,
visible in the colormap of Figure \ref{fig:color} and explicitly in
Figure \ref{fig:profs}, which shows photometric profiles along an
elliptical wedge which follows the curvature of the outermost spiral
arm westward into the outer disk. We see a dramatic drop in surface
brightness between 220\arcsec\ and 250\arcsec, corresponding precisely
with a change in the \bmv\ color from 0.46 to 0.65 \citep[roughly, the
  integrated colors of an Sc and Sa galaxy,
  respectively;][]{roberts94}. This region marks the edge of the
outermost spiral arm; beyond, the redward gradient continues,
ultimately reaching colors similar to those of M51b. The dramatic drop
in surface brightness is similar to the arm-interarm contrast measured
in the same region in M51a by \citet{schweizer76}, and seen in the
outer regions of other spiral galaxies \citep{schweizer76, tacconi90}.

The red colors, smooth surface brightness profile, and round outer
isophotes in this region might be interpreted as marking the
transition to a kinematically hot spheroidal component, but we reject
this notion for a variety of reasons. M51a lacks a strong bulge
  \citep[(B:T)$_I$ = 0.27;][]{pompei97}, arguing that the outer
component is not likely to be extended bulge light. Furthermore, such
a bright halo component is unlikely given the radius ($\sim$ 10 kpc)
and surface brightness ($\mu_{B} \approx 25$) of the component; by
comparison, M31's smooth halo has the much fainter surface brightness
of $\mu_{B}\sim29$--$30$ at 10kpc \citep[extrapolating from the power
  law fits of][]{ibata14}. The most reasonable interpretation for this
component is that we are simply seeing the outer, redder stellar disk
in M51a.

An outer disk origin for this light is also suggested by its
similarity in color to the more extended Northwest Plume. The linear
structure of the Plume would require a dynamical origin from material
with coherent motion and high angular momentum, as would be found in
an outer disk.  Such disks are also expected to show redder colors due
to effects from radial migration \citep[e.g.][]{roskar08}, and such
colors are frequently observed in outer disks \citep[e.g.][who
    found typical outer disk colors of $g-r \sim 0.5$--$0.6 =
    0.7$--$0.8$ in $B-V$]{bakos08}. Hence, it seems likely that the
red outer regions of M51a are simply part of its outer disk, and that
the Northwest Plume may have been ejected from the outer disk during
the interaction.

\subsection{Comparison to simulations}

Simulations of the M51 system fall broadly into two classes: parabolic
single-pass models with typical timescales of $>300$ Myr
\citep{toomre72, hernquist90, durrell03}, and more tightly bound,
multiple passage models with one encounter several hundred Myr ago,
and a second passage $\lesssim$100 Myr ago \citep{howard90, salo00,
  dobbs10}. In general, these models were developed to match the
structure and kinematics of the \ion{H}{1} and the bright tidal
debris (the East and West Streams and Southern Tail). The degree to
which they successfully reproduce the fainter, more extended debris
(the Northwest, Northeast, and South Plumes) is less clear. Here we
develop new insights gleaned from a comparison between our deep
imaging data and the existing simulations.

Perhaps most useful is a comparison involving the Northwest Plume.
Similar features arise in simulations of M51 from the material nearest
the companion at pericenter, which receives the most dispersive tidal
kick and is ejected as a broad tidal plume (in contrast, material on
the opposite side of the disk forms a coherent tidal tail). Tidal
features resembling the Northwest Plume appear in a variety of M51
simulations, but single-passage models tend to reproduce the Plume's
morphology better than the best-fit multipassage models of
\citet{salo00} or \citet{dobbs10}: see, for example, Figures 1 and 8
of \citep{salo00, howard90}, respectively. However, we note that both
\citet{salo00} and \citet{dobbs10} elaborate on only one such model;
different parameters in a multi-passage encounter may reproduce the
feature more effectively.  While the Northwest Plume cannot
unambiguously differentiate between the model classes, the dynamical
age implied by its linear extent argues that the initial passage of
this encounter (or the sole passage) must have taken place at least a
few hundred Myr ago \citep[in more agreement with][]{howard90}.

Unfortunately, the other faint features---the Northeast and South
Plumes---have no clear analogues in any published simulation. The
Northeast Plume perhaps most resembles the diffuse northern component
seen in the multiple passage models by \citet{salo00}. Single passage
models generally completely fail to reproduce this northern component
\citep{salo00, durrell03}, so the presence of both the crown and this
Northeast Plume seem to favor the multiple-passage models. Indeed,
this feature may simply be a faint outer extension of material in the
crown itself; its low luminosity ($\sim 10^6 L_{\ensuremath{\odot},
  B}$) would place it below the mass resolution of most published
simulations, and could explain why they miss this feature.

The South Plume, on the other hand, shows no morphological continuity
with any adjacent feature, either in optical or in \ion{H}{1}, and has no
counterpart in any simulation published to date. While
\citet{appleton86} detected some diffuse \ion{H}{1} on the galaxy's
southwestern side, it does not overlap with the South Plume
itself. Given all this, the Plume would seem to be either a distinct
tidal feature not replicated in previous dynamical modeling (although
like the Northeast Plume, it may be too low mass to discern in
published simulations), or the remains of a disrupted low luminosity
dwarf galaxy in the system.

Regarding the Southeast \ion{H}{1} tail, the lack of an old,
co-spatial stellar component may argue that the gas came from an
extended gas-rich disk belonging to M51a. Simulations of the M51
system favor a highly inclined encounter to explain the kinematics of
the tail \citep{rots90, salo00}; such a configuration would suppress
caustic formation in the tail and explain its paucity of star
formation and young stars. However, the gas/star offset observed in
the tail is still problematic, as such features should be most
prominent in highly prograde encounters where the orbital and
rotational angular momentum is well-aligned \citep{mihos01}. Thus the
data still leave some ambiguity in the dynamical evolution of the
Southeast Tail.

In summary, no single model of M51 provides a perfect match to the
additional constraints derived from our imaging. The single-passage
models appear to better reproduce the diffuse Northwest Plume, while the
multi-passage models reproduce better the diffuse northern material (the
crown and Northeast Plume). No simulations to date have produced
anything resembling the South Plume, although its low luminosity
suggests that higher resolution simulations may be necessary to model
its structure. The lack of a stellar counterpart in M51's Southeast HI
tail argues that the tail was drawn almost exclusively from the
outermost, gas-rich regions of M51a's disk, or that extensive kinematic
segregation of stars and gas has occurred, without triggering any
wide-spread star formation in the tidally stripped gas.

Future modeling campaigns thus are warranted, and have a variety of
new constraints and challenges ahead. New simulations must recover the
newly discovered tidal plumes, ensure that the Northwest Plume is
correctly oriented and gas-free, reproduce the strong offset between
the stars and gas in the Southeast Tail, and bifurcate M51b's East and
West Streams.  Successful simulations will then be able to pin down
the timescale and geometry of the encounter---critical inputs to
studies of spiral structure, ISM evolution, and star formation in this
classic interacting system.

\acknowledgements

The work was supported by the NSF through grant 1108964 to JCM.

\end{document}